# High-order virtual gain for optical loss compensation in plasmonic metamaterials

Fuxin Guan[1,†], Zemeng Lin[1,†], Sixin Chen[1,†], Xinhua Wen[1], and Shuang Zhang[1,2,3,*]

[1]New Cornerstone Science Laboratory, Department of Physics, University of Hong Kong, Hong Kong, China

[2]Department of Electrical & Electronic Engineering, University of Hong Kong, Hong Kong, China

[3]Materials Innovation Institute for Life Sciences and Energy (MILES), HKU-SIRI, Shenzhen, P.R. China

[†] These authors contributed equally this work.

[*] Corresponding author: shuzhang@hku.hk

**Abstract:**

Metamaterials exhibit extraordinary properties yet suffer from pronounced wave dissipation, particularly in optical imaging and sensing systems. Recent advances leveraging complex frequency wave excitations with virtual gain effect, synthesized by multi-monochromatic waves, offer promising solutions for optical loss compensation. However, this approach faces limitations in extreme loss scenarios. The complex frequency wave requires sufficient virtual gain, i.e., temporal attenuation, to offset material loss, inevitably triggering rapid signal decay to zero before reaching a quasi-static state. To address this challenge, we introduce synthetic waves of high-order virtual gain to slow down the decay rate while preserving the loss compensation efficiency. We experimentally demonstrate 20-fold noise suppression in plasmonic resonance systems compared to conventional complex frequency excitations. This approach exhibits broad applicability across diverse fields, including imaging, biosensing, and integrated photonic signal processing.

**Introduction**

Optical losses pose serious constraints on various applications of plasmonics and metamaterials, including superlens imaging [1–6], optical sensing [7–12] and polaritonic information processing [13–17]. Although low-temperature experiments have been proposed as a solution to the material loss, experimental realization is quite challenging despite some progress being made [18]. There have been some endeavors to compensate for optical loss of plasmonics and metamaterials using gain material [19–21]. However, the required gain is unrealistic for many plasmonic systems and gain also introduces instability and noise to the systems.

Complex frequency waves (CFWs) have emerged as critical tools for loss compensation in superlens systems, enabling super-resolution imaging capabilities beyond conventional diffraction limits [5,22–26]. Beyond imaging applications, these engineered waveforms are further driving discoveries of novel phenomena in wave physics [27–34]. However, excitation and measurement involving complex frequency waves with temporally exponential attenuation present challenges in optical experiments. A practical solution has been recently proposed, involving the linear combination of the measurement results at multiple monochromatics in an appropriate manner [6]. This approach has facilitated experimental observations of deep-subwavelength superlens imaging, ultra-sensitive molecular sensing [35], and near loss-free polaritonic propagation [36]. However, the temporally abrupt truncation at the onset of CFWs and the finite response time inherent to realistic systems necessitate a lengthy period to enter quasi- static state, thereby minimizing unwanted noise signals. Systems with extreme losses require a substantial virtual gain, achieved through the implementation of a large imaginary frequency. This requirement induces rapid attenuation of CFW before it stabilizes into a measurable quasi-static state, undermining the efficacy of CFW-based loss compensation. Furthermore, finite available frequency range in measurements also seriously impact the synthesis of complex frequency results.

In this work, we address these limitations by engineering high-order virtual gain (HVG) excitations via the multi-monochromatic synthesis approach, which corresponds to a high-power Lorentzian spectral profile in the frequency domain. It is worth noting that high-power virtual loss has recently been harnessed for excitation of absorbing exceptional point [37–39]. Temporally, the synthetic HVG waveform manifests as an exponential attenuation function modulated by a temporal power-law component, dramatically mitigating the temporal attenuation. Compared to CFWs with the equivalent virtual gains, HVG excitations demonstrate enhanced robustness against the limitation imposed by the rapid temporal attenuation. To demonstrate the concept, we implement synthesized HVG excitations in probing plasmon-induced transparency within optical metamaterials. Our results reveal that the HVG excitation achieves 30-fold greater noise suppresion compared to conventional CFWs.

**Theory**

We begin with a brief description of compensating loss in permittivity using synthetic complex frequency excitation [6]. The excitation wave is a truncated CFW, $E(t) = e^{-i\widetilde{\omega}t}\theta(t) = \int 1/[2\pi i(\widetilde{\omega} - \omega')]e^{-i\omega't}d\omega'$, where $\widetilde{\omega} = \omega - i\beta$, and $\beta > 0$ corresponds to the temporal attenuation factor. The corresponding field profile is shown in Fig. 1(a), which takes an attenuating waveform. We consider a permittivity described by Lorentzian form, $\varepsilon_L(\omega) = 1 - \omega_p^2/(\omega^2 + i\omega\gamma - \omega_0^2 - \gamma^2/4)$, where $\gamma$ represents the damping term, $\omega_p$ and $\sqrt{\omega_0^2 + \gamma^2/4}$ are the plasma frequency and resonance frequency, respectively. Using the synthetic complex frequency approach for a loss compensation, the synthesized displacement field under complex frequency excitation can be expressed as a linear combination of multi-frequency response, $D(\widetilde{\omega}, t) = \int_{\omega_0-\omega_\Delta}^{\omega_0+\omega_\Delta} \varepsilon_L(\omega')e^{-i\omega't}/[2\pi i(\widetilde{\omega} - \omega')]\,d\omega'$, where $2\omega_\Delta$ represents the integral range, determined by the experimentally accessible frequency range for measurements. We then divide it into two parts, $D(\widetilde{\omega}, t) = \int_{-\infty}^{\infty} \varepsilon_L(\omega')e^{-i\omega't}/[2\pi i(\widetilde{\omega} - \omega')]\,d\omega' + D_{FFL}(t)$, where $D_{FFL}(t) = -(\int_{\omega_0+\omega_\Delta}^{\infty} + \int_{-\infty}^{\omega_0-\omega_\Delta})\varepsilon_L(\omega')e^{-i\omega't}/[2\pi i(\widetilde{\omega} - \omega')]\,d\omega'$ represents the unwanted signal due to the finite frequency range, and can be approximated as $|D_{FFL}(t)| \approx$

$|\cos(\omega_\Delta t)|(\pi t \omega_\Delta)^{-1}$. Based on the residue theorem, the displacement field can be further expressed as, $D(\tilde{\omega}, t) = \varepsilon_L(\tilde{\omega})e^{-i\tilde{\omega}t} + D_{TIS}(\tilde{\omega}, t) + D_{FFL}(t)$, where $D_{TIS}(\tilde{\omega}, t) = \omega_p^2 e^{-i\tilde{\omega}_0 t}/[2\omega_0(\tilde{\omega} - \tilde{\omega}_0)]$ represents the unwanted signal caused by the abrupt temporal truncation and the resonance of permittivity. The combined term $(|D_{TIS}(\tilde{\omega}, t)| + |D_{FFL}(t)|)$ represents the overall unwanted signal with the multi-frequency approach. In practice, it is important to minimize the impact of this term by optimizing the evolution period for taking the measurements.

The temporal evolutions of the above three terms are given by: $|\varepsilon_L(\tilde{\omega})e^{-i\tilde{\omega}t}| \propto e^{-\beta t}$ (the target), $|D_{TIS}(\tilde{\omega}, t)| \propto e^{-\gamma t/2}$ and $|D_{FFL}(t)| \propto (\omega_\Delta t)^{-1}$. To illustrate the optimization of evolution period, we plot in Fig. 1(b) the temporal evolution of the three terms in the logarithmic scale for a given set of parameters shown in the figure caption. To reach the quasi-steady state, it is required that $\beta < \gamma/2$ such that $D_{TIS}$ attenuates faster than the target signal, and their ratio is represented by difference between the solid and dashed lines of the same color in Fig. 1(b), which increases linearly with time. On the other hand, the finite frequency effect, represented by the black solid line, remains the same for different virtual gains, and decays at a slower rate than TIS and surpasses it after a certain period. Thus, the optimal evolution time is the moment when $|D_{TIS}(\tilde{\omega}, t)| = |D_{FFL}(t)|$, corresponding to the crossing points between the dashed lines and the black line in Fig. 1(b), which critically depends on the permittivity damping (the $\gamma/2$ term) and the available frequency range $\omega_\Delta$. When the virtual gain is increased (see the blue and green lines), the ratio between the target signal and $D_{TIS}$ at the optimal evolution time becomes lower, leading to a reduction of TUR (target to unwanted signal ratio), defined as ratio between the target signal to the overall unwanted signal (represented by the length of the vertical grey lines in Fig. 1(b)). When $\beta$ approaches $\gamma/2$ as shown by the case of $\beta = 0.45\gamma$ in Fig. 1(b), the ratio is close to unity, making loss compensation almost impossible to reach.

The imaginary parts of the compensated permittivities [$Im(\varepsilon)$] for the three different virtual gains at their respective crossing points are shown in Fig. 1(c). $Im(\varepsilon)$ with higher imaginary frequency $\beta$ (virtual gain) exhibits stronger oscillatory features due to the significant level of the unwanted signals. Thus, there exists a trade-off between the level of virtual gain and the TUR. More details can be found in supplementary section 1.

Next, we show how this tradeoff is alleviated by employing excitations with high-order virtual gain. The $n^{\text{th}}$ power synthetic wave takes the form $E_n(t) = t^{n-1}e^{-i\tilde{\omega}t}\theta(t)$, which can be decomposed into a Fourier form as $E_n(t) = (n-1)!\int e^{-i\omega' t}/[2\pi(i\tilde{\omega}-i\omega')^n]\,d\omega'$, where $\tilde{\omega} = \omega - i\beta$. Note that for $n=1$, $E_1(t) = e^{-i\tilde{\omega}t}\theta(t)$ recovers the truncated CFW. The corresponding waveforms for $n = 2, 3, 4$ are presented in Fig. 2(a). The synthesized response of the $n^{\text{th}}$ order displacement field can be derived as, $D_n(\tilde{\omega}, t) = (n-1)!\int_{-\infty}^{\infty}\varepsilon_L(\omega')e^{-i\omega' t}/[2\pi(i\tilde{\omega}-i\omega')^n]\,d\omega' + D_{n,FFL}(t)$, where $|D_{n,FFL}(t)| = |(n-1)!(\int_{\omega_0+\omega_\Delta}^{\infty} + \int_{-\infty}^{\omega_0-\omega_\Delta})\varepsilon_L(\omega')e^{-i\omega' t}/[2\pi(i\tilde{\omega}-i\omega')^n]\,d\omega'| \approx$

$$\begin{cases}(n-1)!|\cos(\omega_\Delta t)|(\pi t\omega_\Delta^n)^{-1} & n \in odd \\ (n-1)!|\sin(\omega_\Delta t)|(\pi t\omega_\Delta^n)^{-1} & n \in even\end{cases}.$$

Using high-order residue theorem, we have $D_n(\tilde{\omega}, t) \approx t^{n-1}\varepsilon_L(\tilde{\omega})e^{-i\tilde{\omega}t} + D_{n,TIS}(\tilde{\omega}, t) + D_{n,FFL}(t)$, where $D_{n,TIS}(\tilde{\omega}, t) = i(n-1)!\omega_p^2 e^{-i\tilde{\omega}_0 t}/[2\omega_0(i\tilde{\omega}-i\tilde{\omega}_0)^n]$. The temporal evolutions of the three terms are given by: $|t^{n-1}\varepsilon_L(\tilde{\omega})e^{-i\tilde{\omega}t}| \propto e^{-\beta t}t^{n-1}$, $|D_{n,TIS}| \propto e^{-\gamma t/2}$ and $|D_{n,FFL}| \propto (\omega_\Delta^n t)^{-1}$. The evolutions of the target and unwanted signals for HVGs of different orders are given in Fig. 2(b) with case I: $n=1$, $\beta = 0.2\gamma$ (red); case II: $n=2$, $\beta = 0.4\gamma$ (blue); case III: $n=3$, $\beta = 0.492\gamma$ (green). Here we choose different virtual gains for different orders to ensure that they have the same TUR around 1% at their respective optimized evolution moments, where the $n^{\text{th}}$ displacement field can be approximated as, $D_n(\tilde{\omega}, t) \approx t^{n-1}\varepsilon_L(\tilde{\omega})e^{-i\tilde{\omega}t}$ (supplementary section 2-3). The multiplication of $t^{n-1}$ dramatically slows down the attenuation of the excitation, making displacement field of high orders more robust to the unwanted signals arising from temporal truncation and finite frequency range than that of complex frequency excitation. The corresponding real and imaginary parts of the compensated permittivities for $n = 1$ to $4$, together with that without loss compensation, are shown in Fig. 2(c) and Fig 2(d), respectively. The corresponding full width at half maximum (FWHM) of the imaginary permittivity $\Delta\omega_{fw}$ for $n = 0$ to 4 ($n = 0$ represents the case without loss compensation) are shown in the inset of Fig. 2(d). It shows that by maintaining the same TUR level, the HVGs can lead to more effective loss compensation. For the particular set of parameters, the loss compensation starts to saturate around $n = 3$. More detailed information for the temporal evolution of the permittivity with virtual gains of different orders are presented in supplementary Figs. S1 and S2.

It is important to note that, although the above analysis is applied to the permittivity, any linear responses of the system under excitation of $n^{th}$ order HVG can be synthesized via the multi-frequency approach, given by

$$F_n(\widetilde{\omega}, t) \approx (n-1)! \sum_k \frac{F(\omega'_k) e^{-i\omega'_k t} \Delta\omega'_k}{2\pi(i\widetilde{\omega} - i\omega'_k)^n} \tag{1}$$

where $\Delta\omega'_k$ is the frequency interval.

**Loss compensation for plasmon-induced transparency**

We utilize a plasmon-induced transparency (PIT) system to further investigate the effectiveness of loss compensation using HVG. Optical sensors based on plasmonic/phononic nanostructures have been extensively investigated [40,41]. However, optical losses in the nanostructures dramatically reduce their sensitivity. Specifically, for a PIT system [42,43], plasmonic losses reduce the quality factor of the subradiant mode, leading to dramatically degraded transparency peak, which is typically used for investigation of spectroscopy. To address this issue, we employ the synthetic excitation of HVG to recover the transparency peak in the measurements. The PIT metamaterial is designed following Ref. [43], where each dolmen shaped unit cell consists of coupled bright and dark elements, supporting a radiative mode and a subradiant mode, respectively, as shown in Fig. 3(a). The subradiant mode is supported by two parallel metal bars (enclosed in the dashed box), while the radiative mode is supported by the vertical metal bar. When a vertically polarized light shines on the metamaterial, it excites the radiative mode, which then couples to the subradiant mode, forming a PIT system. The nanofabrication and measurement techniques are displayed in the method part of supplementary.

The measured transmission ($T_M$) spectrum of the PIT sample is presented in Fig. 3(b) as the black curve. Due to the existence of significant plasmonic loss, the transparency peak at around 2850 cm$^{-1}$ is barely visible. To compensate the plasmonic loss using the synthesized HVG, an intermediate polarization parameter $\tilde{P}(\omega) = i(\frac{1}{\tilde{t}_M(\omega)} - \frac{1}{t_0})$ is employed, similar to the approach employed in Ref. [35], where $\tilde{t}_M(\omega) = e^{i\varphi_M}\sqrt{T_M}$ corresponds to the transmission coefficient, and the phase term $\varphi_M$ is obtained using the Kramers-Kronig relations [44]. The Parameter $t_0 \approx 2/(1 + n_s)$ denotes the transmission coefficient of the pure substrate [45], where $n_s \approx 1.35$ corresponds to the refractive index of substrate made of fused silica. The magnitude of the calculated polarization term $\tilde{P}(\omega)$ is shown in Fig. 3(c) as the black line. We input $\tilde{P}(\omega)$ into Eq. (1) to obtain $\tilde{P}_n(\tilde{\omega}, t)$ of different orders, and the corresponding temporal evolution of their magnitudes are shown in Fig. 3(d) for $n = 1, 2, 3$, with the same virtual gain of $200\ cm^{-1}$. The polarization $\tilde{P}_1(\tilde{\omega}, t)$ exhibits quite significant unwanted oscillatory features over time and frequency, while the oscillatory features are very small for $\tilde{P}_2(\tilde{\omega}, t)$ and $\tilde{P}_3(\tilde{\omega}, t)$. For these cases, we select the temporal snapshots at white dashed lines in Fig. 3(d) to obtain the compensated polarizations, which are plotted as the red lines in Fig. 3(c). The corresponding transmittances $|\tilde{T}_n(\tilde{\omega}, t)|$ are plotted in Fig. 3(b) as the red lines. Although the excitation of $n = 1$ can recover the PIT peak in the transmission spectrum, the oscillatory features dramatically reduce the TUR. On the other hand, the HVG excitations of $n = 2$ and 3 have greatly enhanced the TUR of the transmittance, exhibiting loss compensated spectra with good quality. Additionally, the other weaker coupling case is shown in Fig. S3 for a better comparison.

**Summary**

We have proposed synthetic waves of high-order virtual gain to compensate for the intrinsic losses of plasmonic metamaterials, achieving 20-fold noise suppression in transmission spectra, which surpasses the performance attainable with conventional complex frequency excitation. Our approach effectively resolves the spectral oscillations induced by temporal truncation and finite frequency limitations in synthesized complex frequency excitations, demonstrating its potential for high-precision spectroscopy and molecule sensing. This approach is broadly applicable to loss mitigation in plasmonic, phononic, and excitonic systems, enabling long-distance polaritionic propagations and facilitating advancements in on-chip photonics. Moreover, it can be extended to other wave regimes, including acoustic and hydrodynamic waves.

# Acknowledgments

**Funding:** This work was supported by the New Cornerstone Science Foundation, the Research Grants Council of Hong Kong (AoE/P-502/20, STG3/E-704/23-N, 17309021), Guangdong Provincial Quantum Science Strategic Initiative (GDZX2204004, GDZX2304001).

**Author contributions:** S.Z. and F.G. conceived the project and supervised the overall projects. Z.L. fabricated the PIT samples and carried out the FTIR experiment. F.G. and S.C. performed numerical simulations and analytical calculations. F.G., Z.L., S.C., K.Z., X.W., and S.Z. participated in the analysis of the results. F.G. and S.Z. wrote the manuscript with input from all authors. All authors contributed to the discussion.

**Competing interests:** The authors declare no conflicts of interest.

**Data and materials availability:** All data are available in the main text or the supplementary materials.

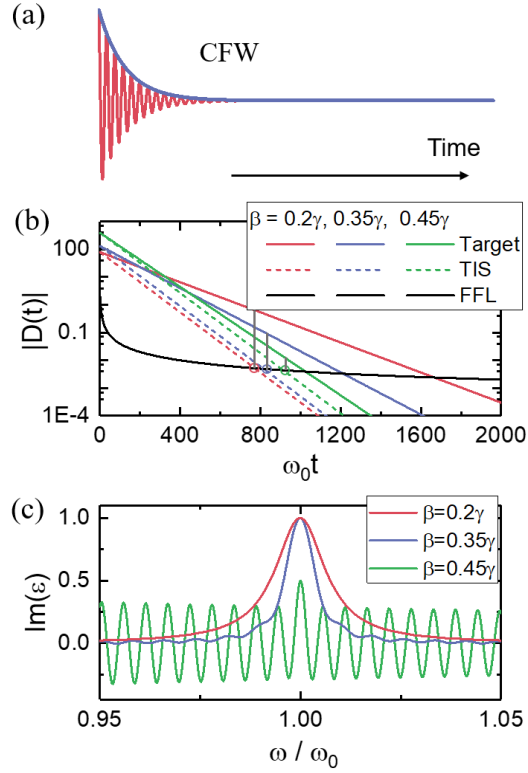

Figure. 1. Illustration of loss compensation with CFW of different virtual gains. (a) Temporal profile of a truncated complex frequency wave. (b) Temporal evolutions of displacement fields of target (colored solid lines), TIS (colored dashed lines) and FFL (black line) under synthesized complex frequency excitation. The red, blue and green lines correspond to different virtual gains of $\beta = 0.2\gamma$, $0.35\gamma$, and $0.45\gamma$, respectively. The crossing points between TIS (dashed lines) and FFL (black line) are marked by circles of different colors. The parameters of the Lorentz model are $\omega_p = \omega_0 = 2$, $\gamma = 0.05$. The values of the three TURs are represented by the lengths of grey vertical lines. (c) The imaginary parts of the compensated permittivity $[Im(\varepsilon)]$ with the three different $\beta$ values at the crossing points in (b). The integral range is $(0.5\omega_0 \sim 1.5\omega_0)$.

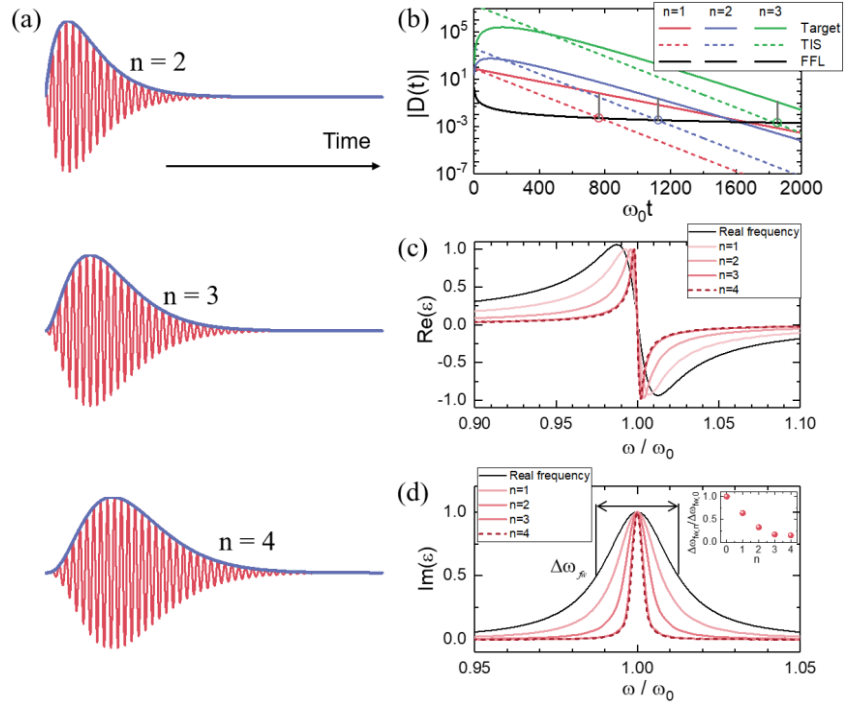

Figure. 2. Synthetic excitations of HVGs for loss compensation. (a) Temporal profiles of HVG waves. (b) Temporal evolutions of target (colored solid lines), TIS (colored dashed lines) and FFL (black line) under excitation of HVGs of three different orders, I: $n$=1, $\beta = 0.2\gamma$; II: $n$=2, $\beta = 0.4\gamma$; III: $n$=3, $\beta = 0.492\gamma$. Here the FFL terms in three cases are normalized to the same curve for better comparison. (c) Real and (d) imaginary parts of the Lorentz permittivity with the synthesized excitations of high-order virtual gains. The virtual gains of different orders, ranging from $n$=1 to 4, are given by $\beta_1 = 0.2\gamma$, $\beta_2 = 0.4\gamma$, $\beta_3 = 0.492\gamma$, $\beta_4 = 0.499\gamma$, respectively. Each result is selected at the snapshot of crossing point in (b). The inset shows the dependence of FWHM over the orders of different virtual gains.

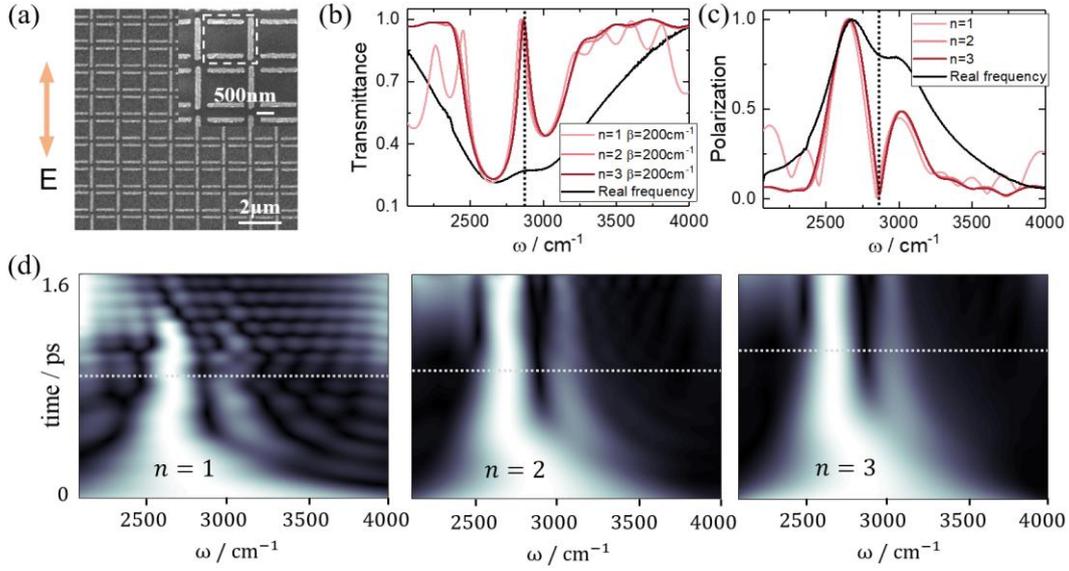

Figure. 3. Experimental demonstration of recovering plasmonic resonances with HVG excitations. (a) Scanning electron microscopy (SEM) image of the fabricated PIT metamaterial. (b) Transmittances ($\tilde{T}$ and $\tilde{T}_n$) and (c) polarizations ($\tilde{P}$ and $\tilde{P}_n$) under excitations of real frequency and HVGs. The same virtual gain of $\beta = 200$ cm$^{-1}$ is used for all cases. The vertical dashed line indicates the simulated peak position. (d) The temporal evolutions of polarizations $\tilde{P}_n(\tilde{\omega}, t)$ for $n = 1, 2,$ and $3$. The spectra shown in (c) are selected at the moments indicated by the dashed lines in (d).

**References**:

[1] J. B. Pendry, Negative refraction makes a perfect lens, Phys Rev Lett **85**, 3966 (2000).

[2] N. Fang, H. Lee, C. Sun, and X. Zhang, Sub–Diffraction-Limited Optical Imaging with a Silver Superlens, Science (1979) **308**, 534 (2005).

[3] Z. Liu, H. Lee, Y. Xiong, C. Sun, and X. Zhang, Far-field optical hyperlens magnifying sub-diffraction-limited objects, Science (1979) **315**, 1686 (2007).

[4] T. Taubner, D. Korobkin, Y. Urzhumov, G. Shvets, and R. Hillenbrand, Near-field microscopy through a SiC superlens, Science (1979) **313**, 1595 (2006).

[5] S. Kim, Y. Peng, S. Yves, and A. Alù, Loss Compensation and Super-Resolution with Excitations at Complex Frequencies, Phys Rev X **13**, 041024 (2023).

[6] F. Guan, X. Guo, K. Zeng, S. Zhang, Z. Nie, S. Ma, Q. Dai, J. Pendry, X. Zhang, and S. Zhang, Overcoming Losses in Superlenses with Synthetic Waves of Complex Frequency, 2023.

[7] N. W. Bartlett, M. T. Tolley, J. T. B. Overvelde, J. C. Weaver, B. Mosadegh, K. Bertoldi, G. M. Whitesides, and R. J. Wood, A 3D-printed, functionally graded soft robot powered by combustion, Science (1979) **349**, 161 (2015).

[8] N. Liu, L. Langguth, T. Weiss, J. Kästel, M. Fleischhauer, T. Pfau, and H. Giessen, Plasmonic analogue of electromagnetically induced transparency at the Drude damping limit, Nat Mater **8**, 758 (2009).

[9] N. Liu, M. Mesch, T. Weiss, M. Hentschel, and H. Giessen, Infrared perfect absorber and its application as plasmonic sensor, Nano Lett **10**, 2342 (2010).

[10] D. Rodrigo, O. Limaj, D. Janner, D. Etezadi, F. J. García De Abajo, V. Pruneri, and H. Altug, Mid-infrared plasmonic biosensing with graphene, Science (1979) **349**, 165 (2015).

[11] K. V. Sreekanth, Y. Alapan, M. Elkabbash, E. Ilker, M. Hinczewski, U. A. Gurkan, A. De Luca, and G. Strangi, Extreme sensitivity biosensing platform based on hyperbolic metamaterials, Nat Mater **15**, 621 (2016).

[12] A. Tittl, A. Leitis, M. Liu, F. Yesilkoy, D.-Y. Choi, D. N. Neshev, Y. S. Kivshar, and H. Altug, Imaging-Based Molecular Barcoding with Pixelated Dielectric Metasurfaces Downloaded From, 2018.

[13] D. N. Basov, M. M. Fogler, and F. J. García De Abajo, *Polaritons in van Der Waals Materials*, Science.

[14] S. I. Bozhevolnyi, V. S. Volkov, E. Devaux, J.-Y. Laluet, and T. W. Ebbesen, Channel plasmon subwavelength waveguide components including interferometers and ring resonators, Nature **440**, 508 (2006).

[15] R. F. Oulton, V. J. Sorger, T. Zentgraf, R. M. Ma, C. Gladden, L. Dai, G. Bartal, and X. Zhang, Plasmon lasers at deep subwavelength scale, Nature **461**, 629 (2009).

[16] Z. Fei et al., Gate-tuning of graphene plasmons revealed by infrared nano-imaging, Nature **486**, 82 (2012).

[17] H. Hu et al., Gate-Tunable Negative Refraction of Mid-Infrared Polaritons, 2023.

[18] G. X. Ni et al., Fundamental limits to graphene plasmonics, Nature **557**, 530 (2018).

[19] S. Xiao, V. P. Drachev, A. V. Kildishev, X. Ni, U. K. Chettiar, H. K. Yuan, and V. M. Shalaev, Loss-free and active optical negative-index metamaterials, Nature **466**, 735 (2010).

[20] J. M. Hamm, S. Wuestner, K. L. Tsakmakidis, and O. Hess, Theory of light amplification in active fishnet metamaterials, Phys Rev Lett **107**, 1 (2011).

[21] M. Sadatgol, Ş. K. Özdemir, L. Yang, and D. O. Güney, Plasmon Injection to Compensate and Control Losses in Negative Index Metamaterials, Phys Rev Lett **115**, 035502 (2015).

[22] A. Archambault, M. Besbes, and J. J. Greffet, Superlens in the time domain, Phys Rev Lett **109**, 097405 (2012).

[23] A. Ghoshroy, Ş. K. Özdemir, and D. Ö. Güney, Loss compensation in metamaterials and plasmonics with virtual gain [Invited], Opt Mater Express **10**, 1862 (2020).

[24] H. S. Tetikol and M. I. Aksun, Enhancement of Resolution and Propagation Length by Sources with Temporal Decay in Plasmonic Devices, Plasmonics **15**, 2137 (2020).

[25] W. K. C. W. M. P. S. W. X. L. X. M. X. L. Yi Yang, Enhancing imaging resolution of superlens through transient illumination, Advanced Fiber Laser Conference **13104**, 718 (2024).

[26] S. An, T. Liu, J. Zhu, and L. Cheng, Complex-frequency calculation in acoustics with real-frequency solvers, Phys Rev B **111**, (2025).

[27] D. G. Baranov, A. Krasnok, and A. Alù, Coherent virtual absorption based on complex zero excitation for ideal light capturing, Optica **4**, 1457 (2017).

[28] H. Li, A. Mekawy, A. Krasnok, and A. Alù, Virtual Parity-Time Symmetry, Phys Rev Lett **124**, 193901 (2020).

[29] G. Trainiti, Y. Radi, M. Ruzzene, and A. Alù, Coherent virtual absorption of elastodynamic waves, Sci Adv **5**, 1 (2019).

[30] S. Kim, S. Lepeshov, A. Krasnok, and A. Alù, Beyond Bounds on Light Scattering with Complex Frequency Excitations, Phys Rev Lett **129**, 203601 (2022).

[31] Z. Gu, H. Gao, H. Xue, J. Li, Z. Su, and J. Zhu, Transient non-Hermitian skin effect, Nat Commun **13**, 7668 (2022).

[32] J. Hinney, S. Kim, G. J. K. Flatt, I. Datta, A. Alù, and M. Lipson, Efficient excitation and control of integrated photonic circuits with virtual critical coupling, Nat Commun **15**, 2741 (2024).

[33] J. Hinney, S. Kim, G. J. K. Flatt, I. Datta, A. Alù, and M. Lipson, Efficient excitation and control of integrated photonic circuits with virtual critical coupling, Nat Commun **15**, (2024).

[34] S. Kim, A. Krasnok, and A. Alù, *Complex-Frequency Excitations in Photonics and Wave Physics*, Science (New York, N.Y.).

[35] K. Zeng, C. Wu, X. Guo, F. Guan, Y. Duan, and L. Lauren, Synthesized complex-frequency excitation for ultrasensitive molecular sensing, ELight **4**, 1 (2023).

[36] F. Guan et al., Compensating losses in polariton propagation with synthesized complex frequency excitation, Nat Mater **23**, 506 (2024).

[37] A. Farhi, A. Mekawy, A. Alù, and D. Stone, Excitation of absorbing exceptional points in the time domain, Phys Rev A (Coll Park) **106**, 1 (2022).

[38] A. Farhi, A. Cerjan, and A. D. Stone, Generating and processing optical waveforms using spectral singularities, Phys Rev A (Coll Park) **109**, 1 (2024).

[39] A. Farhi, W. Dai, S. Kim, A. Alù, and D. Stone, Efficient general waveform catching by a cavity at an absorbing exceptional point, Phys Rev A (Coll Park) **109**, L041502 (2024).

[40]   A. Tittl, A. Leitis, M. Liu, F. Yesilkoy, D. Y. Choi, D. N. Neshev, Y. S. Kivshar, and H. Altug, Imaging-based molecular barcoding with pixelated dielectric metasurfaces, Science (1979) **360**, 1105 (2018).

[41]   D. N. Basov, M. M. Fogler, and F. J. García De Abajo, Polaritons in van der Waals materials, Science (1979) **354**, 195 (2016).

[42]   M. Fleischhauer, A. Imamoglu, and P. J. Marangos, Electromagnetically induced transparency, Rev Mod Phys **77**, 633 (2005).

[43]   S. Zhang, D. A. Genov, Y. Wang, M. Liu, and X. Zhang, Plasmon-induced transparency in metamaterials, Phys Rev Lett **101**, (2008).

[44]   B. Gralak, M. Lequime, M. Zerrad, and C. Amra, Phase retrieval of reflection and transmission coefficients from Kramers–Kronig relations, Journal of the Optical Society of America A **32**, 456 (2015).

[45]   G. Zheng, H. Mühlenbernd, M. Kenney, G. Li, T. Zentgraf, and S. Zhang, Metasurface holograms reaching 80% efficiency, Nat Nanotechnol **10**, 308 (2015).